\documentclass[a4paper,11pt]{article}
\usepackage{times,cite}
\usepackage{graphicx,epsfig}    % if you need to include figures in PostScript
\usepackage[tbtags]{amsmath}
\usepackage{here}

\begin{document}
\begin{center}

{\LARGE \bf Experimental proton-proton correlation function derived for the $pp \to pp\eta$ reaction}\\
\vspace{0.5cm}
{\large P. Klaja$^{a,b}$, P.~Moskal$^{a,b}$ and A.~Deloff$^{c}$ for the COSY--11 collaboration\\}
\vspace{0.5cm}
{\small \it
$^a$Nuclear Physics Department, Jagellonian University, 30-059 Cracow, Poland\\
$^b$Institut f\"ur Kernphysik, Forschungszentrum J\"ulich, 52425 J\"ulich, Germany\\
$^c$Institute for Nuclear Studies, Warsaw, Poland}\\
\end{center}
\vspace{0.5cm}
{\small {\bf Abstract.} Based on the high statistics data 
from the $pp\to ppX$ reaction measured 
by the COSY-11 collaboration~\cite{prc69} we have derived a two-proton correlation function for the production of the $pp\eta$ and $pp+pions$ systems.
 The measured correlation function normalized to the value 
 simulated for a point-like source was compared
 with a theoretical prediction in order
 to estimate the size of the reaction volume.}

\section*{\normalsize INTRODUCTION}

The momentum correlations of particles at small relative velocities are 
widely used to study the spatio-temporal characteristics of
 the production processes
in the relativistic heavy ion collisions~\cite{lisa}.
This technique, called after Lednicky {\em a correlation femtoscopy}~\cite{led}, originates from photon intensity interferometry 
initiated by Hanbury-Brown and Twiss~\cite{hbt}. 
Implemented into the nuclear physics~\cite{led, koonin, kopyl} 
it permits to 
determine the duration of the emission process and the sizes
 of the source from which the particles are emitted~\cite{led}.
A central role plays the correlation function
which has been defined as a ratio of the measured
two-particle distribution divided by the  
  reference spectrum obtained from the former by mixing the particles
 from different events~\cite{led}.
 The importance of the correlation femtoscopy 
 has been well established for the investigations of the dynamics in
heavy ion collisions with  high multiplicity.
However, as pointed out by Chajecki~\cite{chaj}, 
in the case of low-multiplicity collisions 
the interpretation of the correlation function 
measurements is still not fully satisfactory,  
especially in view of the surprising STAR
collaboration observation
 indicating universality of the resulting femtoscopic radii 
for both,
the hadronic  (proton-proton), and heavy ion collisions~\cite{chajlis}.
One of the  challenging issues in this context is the understanding of the contributions from the non-femtoscopic correlations 
which may be induced by the decays of resonances,  
global conservations lows~\cite{chaj},
or by other unaccounted for interactions.

 In particle physics the best place to study two-proton correlations 
 is a kinematically complete measurement of meson production 
 in the collisions of hadrons.
 Particularly favourable are exclusive experiments conducted 
 close to the kinematical threshold where
 the fraction of the available phase-space associated
 with low relative momenta between ejectiles is large~\cite{review}.
\par 
 In this note we report on the $\eta$ meson and multi-pion production experiment
 in which the mesons were generated  
 in the collisions of protons at the beam momentum close to the kinematical
 threshold for the $pp\to pp\eta$ reaction.
 The measurement of the two-proton correlation function for these reactions 
 is  important not only in the context of the
 studies of the dynamics underlying the heavy ion physics. 
 Such investigations are interesting by themselves 
 because they offer a new promissing diagnostic tool,
 still not exploited, for studying the dynamics of meson production 
 in the collisions of hadrons. 
\par
 The correlation function carries information about the
 emitting source and, in particular,  
 about of the size of the interaction volume of 
 the $pp\to pp\eta$ process. The knowledge of this size
 might be essential to answer the intriguing question whether    
 the three-body $pp\eta$ system is capable of supporting 
 an unstable Borromean bound state. 
 The  Borromean systems may be realized in the variety
 of objects on the macroscopic (e.g. strips of papers),
 molecular~\cite{chichak, cantrill} 
 and nuclear
 scale (e.g. $^{11}$Li or $^6$He nuclei~\cite{zhukov, marques, bertulani}).
 According to Wycech~\cite{wycech},  the large enhancement
 of the excitation function for the $pp\to pp\eta$ reaction observed 
 close to the kinematical threshold
 may be explained by assuming that the proton-proton pair is emitted 
 from a large (Borromean like) object whose radius is about 4~fm.

\section*{\normalsize TWO-PROTON CORRELATION FUNCTION}

The experiment was conducted using the proton beam of the cooler synchrotron COSY~\cite{cosy} 
and the internal hydrogen cluster target~\cite{domb}.
Momentum vectors of outgoing protons from the $pp\to ppX$ reaction
were measured by means of the COSY-11 facility~\cite{brauksiepe}. 
The two-proton correlation function $R(q)$ 
was determined for the $pp\eta$ and $pp(m\pi)$ systems, respectively. 
It was calculated  
as a ratio of the reaction yield 
$Y(q)$ to the uncorrelated
yield $Y^*(q)$ according to the formula (cf.~\cite{boal})
\begin{equation}
   R(q)+1 = C^*~\frac{Y(q)}{Y^*(q)},
\label{equ:3}
\end{equation}
where $C^*$ denotes an appropriate normalization constant. 
$Y^*(q)$ was derived from the uncorrelated reference sample obtained 
 by using the event mixing technique~\cite{kopyl}.
Here, $R(q)$ denotes 
a projection of the correlation function 
onto the relative momentum of emitted
particles $q = |\vec{p_{1}}-\vec{p_{2}}|$~\footnote{
Note, that some authors instead of $q$ take  as the independent variable 
the proton-proton center-of-mass momentum $k=q/2$
(c.f. the accompanying contribution~\cite{deloff}).  } 

\subsection*{\normalsize Separation of events from the production of $pp\eta$ and $pp+pions$ systems}

In the discussed experiment, only 
four-momenta of two protons were measured and the unobserved meson
was identified via the missing mass technique \cite{prc69, habil}. 
In such situation an entire accessible information about an event is contained
in the momentum vectors of registered protons.
Therefore, it is impossible to know 
whether in a given event the $\eta$ meson or a few pions have been created.
However, statistically, one can separate these groups of events 
on the basis of the missing mass spectra,
for each  chosen region of the phase-space.
\begin{figure}[H]
  \includegraphics[height=.26\textheight]{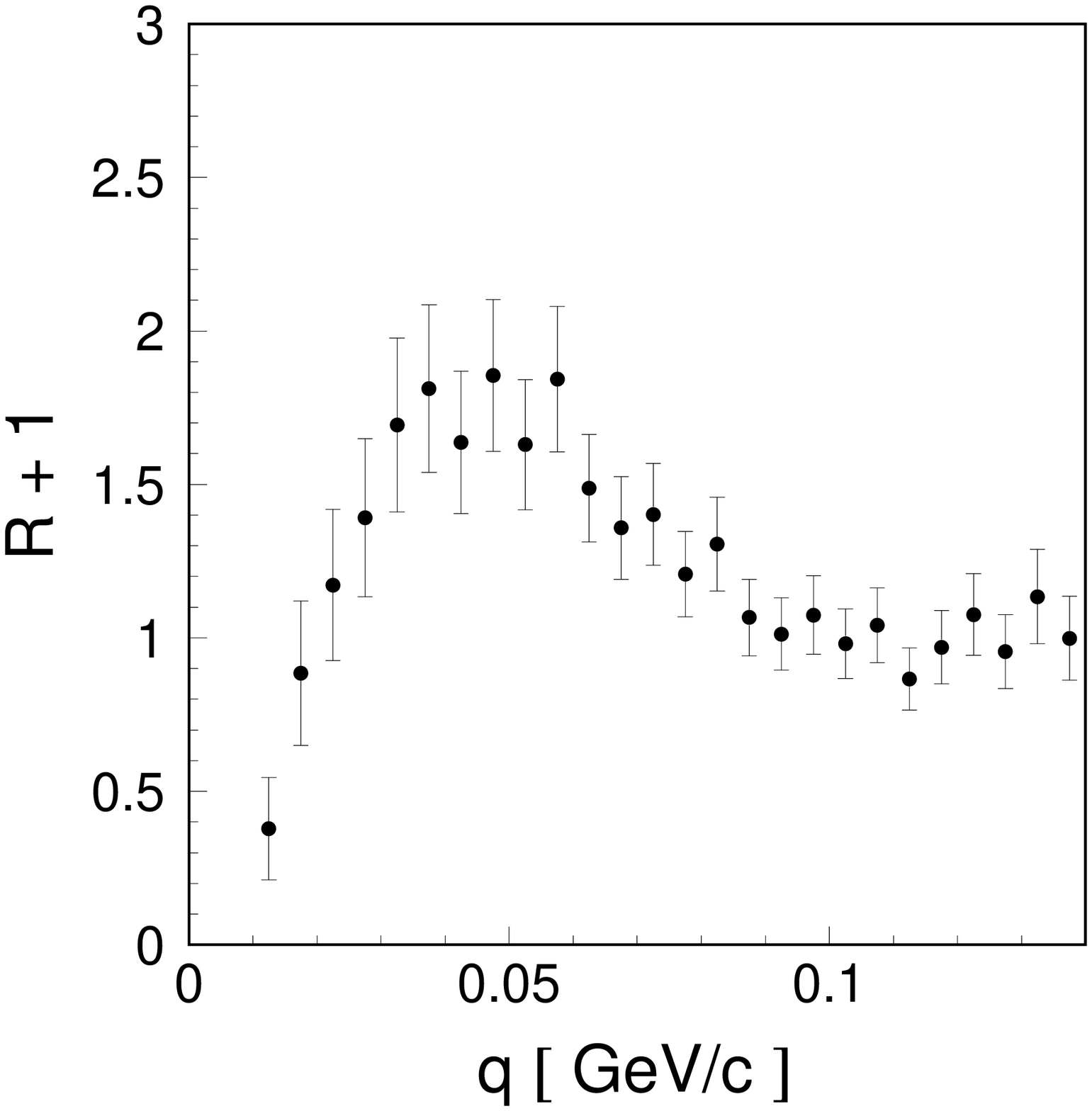}
  \includegraphics[height=.26\textheight]{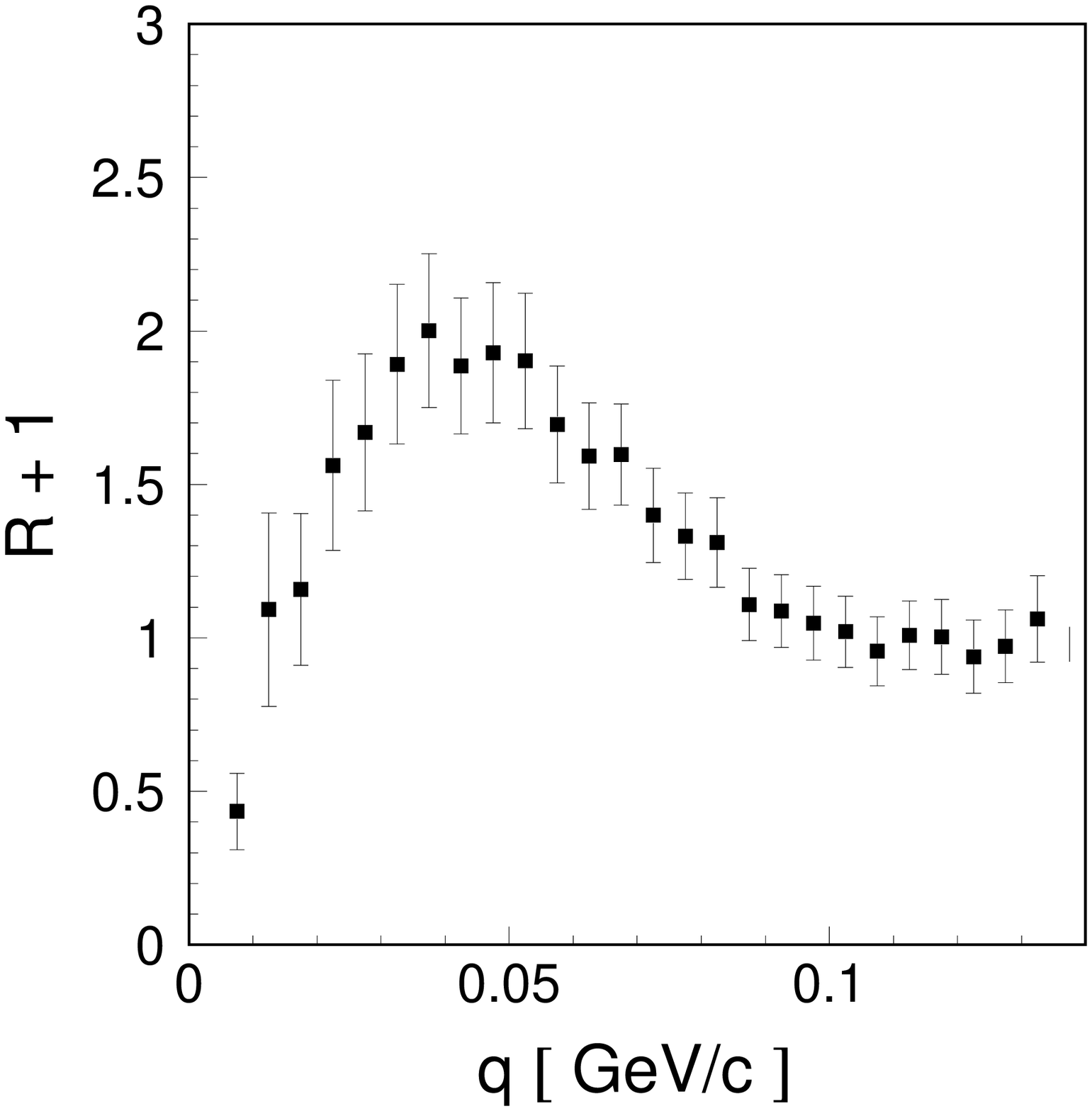}
  \caption{\footnotesize Experimental two-proton correlation function determined 
  for the $pp \to pp\eta$ reaction (left panel), 
  and for  the $pp\to pp+pions$ 
  reaction (right panel). The result shown in this figure is not corrected 
  for the acceptance of the detection system.}
\label{corr:outaccep}
\end{figure}
The derivation of the denominator in (\ref{equ:3}) was much more complicated.
In order to extract $Y^*$ we have developed a method based on the assignment of the statistical weigths 
to the registered events. The weights describe the probability that a given event corresponds to the $pp\eta$
or to the $pp+pions$ production. This technique allowed us for the derivation of the 
uncorrelated yield separately for the production of the $\eta$ meson and for pions.
For details the interested reader is referred to~\cite{acta}.
Figure~\ref{corr:outaccep} presents the determined  correlation functions 
but at this stage of the
evaluation the result is biased by the limited detection acceptance and efficiency
to be considered in the next section.

\subsection*{\normalsize Acceptance corrections}

As the next necessary step in the data evaluation we have corrected the determined yields
to account for the finite geometrical acceptance of the COSY-11 detection setup.

\begin{figure}[H]
  \includegraphics[height=.26\textheight]{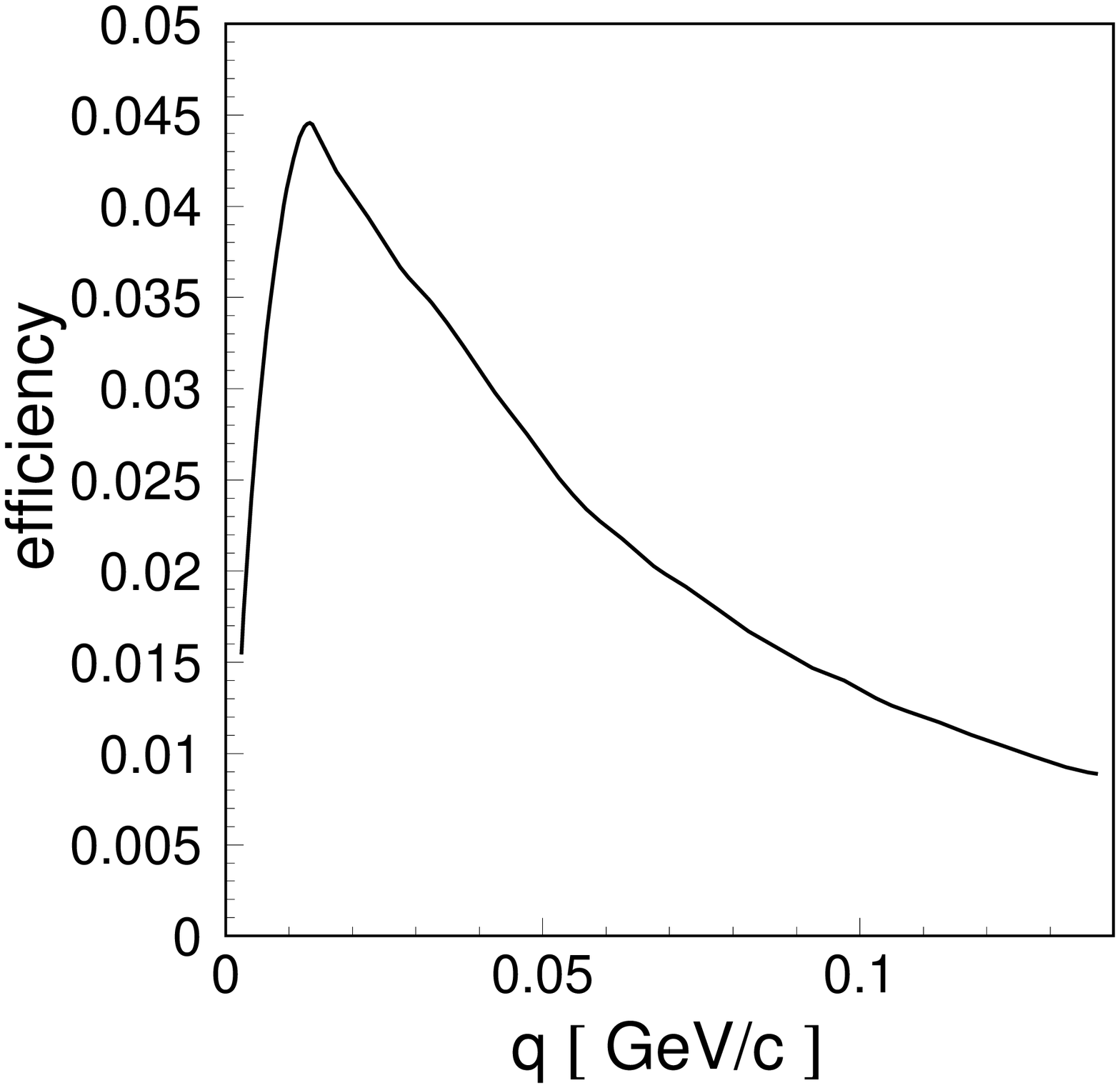}
  \includegraphics[height=.26\textheight]{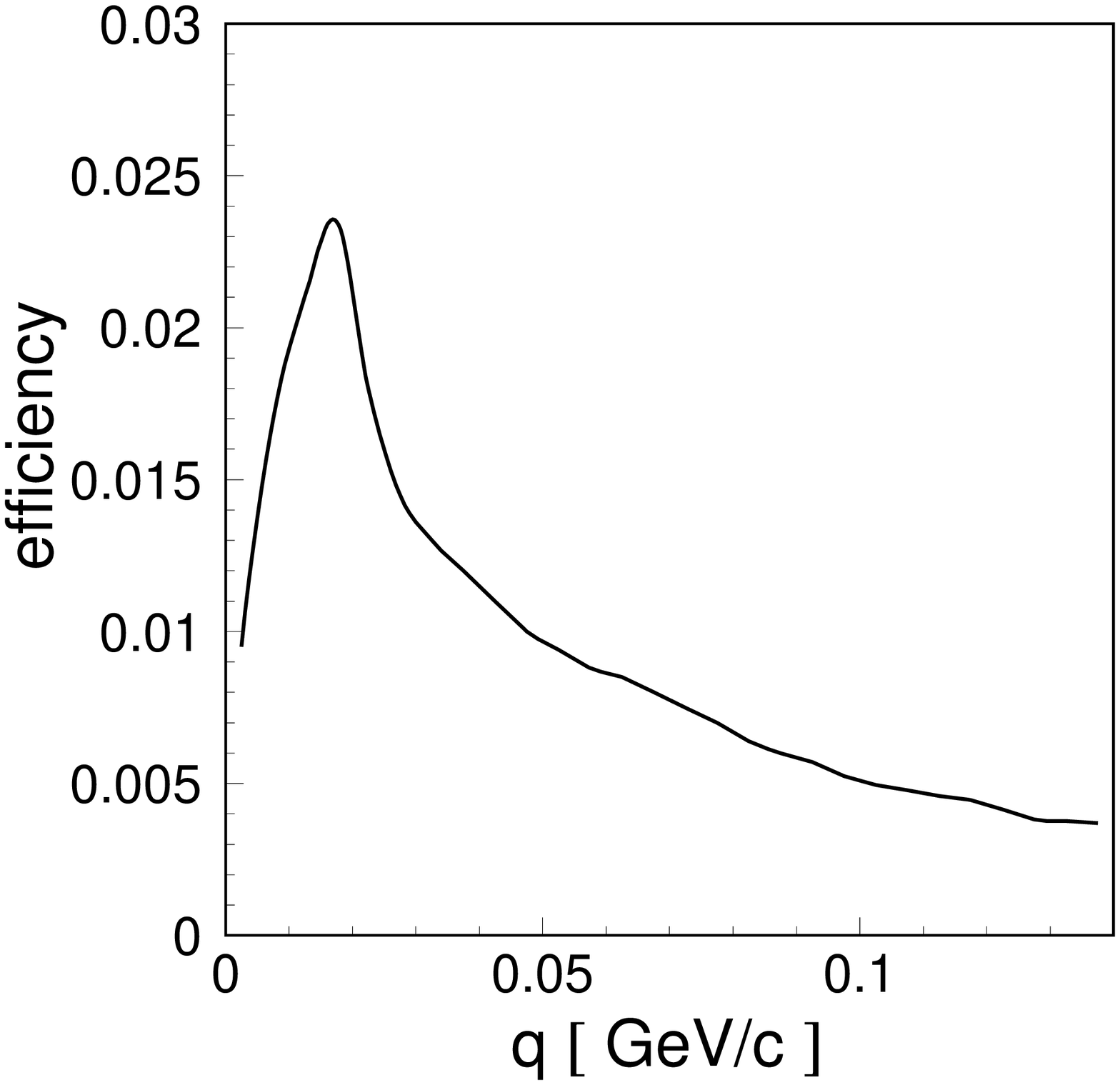}
  \caption{\footnotesize Overall detection efficiency (acceptance and efficiency) of the COSY-11 detection setup
                       for the measurement of the $pp \to pp\eta$ reaction at the excess energy of Q~=~15.5~MeV (left panel) 
                       and 
		       for the measurement of the $pp \to pp(m\pi)$ reactions with the invariant mass of m$\pi$
                       system equal to the mass of the $\eta$ meson
		       (right panel).}
\label{accep}
\end{figure}

First, we 
calculated the acceptances and efficiencies of the COSY-11 system 
for the registration and reconstruction of the $pp\to pp\eta$ and $pp\to pp(m\pi)$ reactions
as  functions of the relative momentum of the outgoing protons. 
The results of the simulations 
are presented in the figure~\ref{accep}.

Knowing the acceptance it would be straightforward to correct the nominator
 in (\ref{equ:3}), however the correction 
of the uncorrelated
yield $Y^*(q)$ is not trivial since the
momenta of the protons in the uncorrelated event originate from 
two independent real events which, in general, could correspond to different
values of the detection efficiency.

Therefore, in order to derive a correlation function 
corrected for the acceptance,  we have created a sample of data
that would have beeen measured with an ideal detector. For this purpose 
we multiplied each reconstructed event so many times 
as it results from the known acceptance.
This means that a given reconstructed 
$pp\to pp\eta$ ($pp \to pp(m\pi)$, respectively) event 
with a proton-proton relative momentum equal q
was added to the sample 1/A(q) times. 

With the aid of this corrected data sample 
we calculated the two-proton correlation function
according to the formula (\ref{equ:3}). 
In order to 
avoid mixing between the same events, a   
'mixing step' in the calculations was set to a value bigger than 
the inverse of  the lowest acceptance value. 
A random repetition of the identical combinations 
was also omitted by increasing correspondingly a 'mixing step'.  
In particular, a k$^{th}$ real event, from the acceptance corrected data sample,
was "mixed" with a (k+n)$^{th}$
event, where $n~>~ max(1/A(q))$. 
If the (k+1)$^{th}$ event was the same as k$^{th}$,
then this was mixed with  a (k+1+2n) event, etc.

\subsection*{\normalsize Results}

\begin{figure}[H]
  \includegraphics[height=.2\textheight]{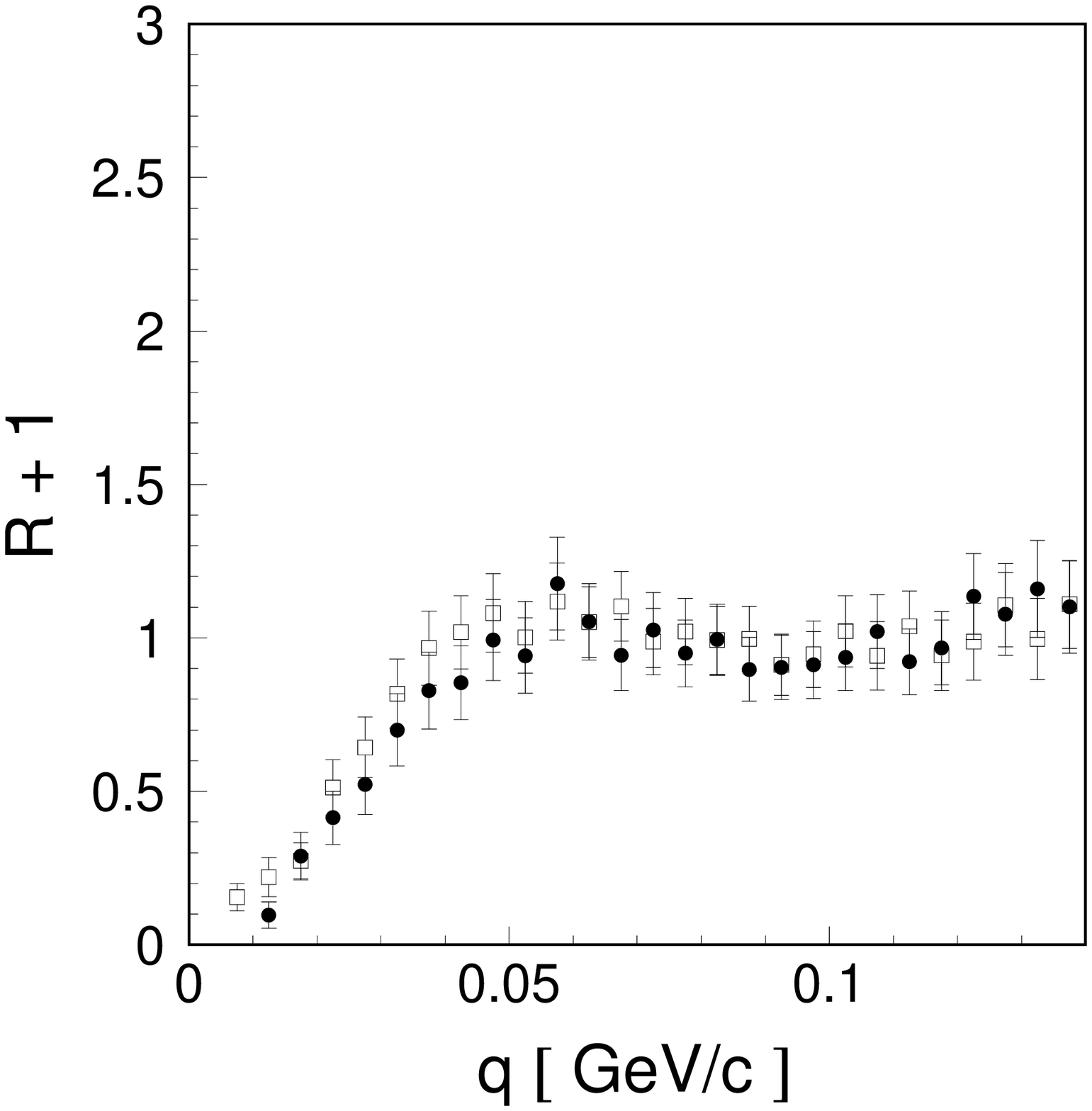}   \hspace{-0.8cm}
  \includegraphics[height=.2\textheight]{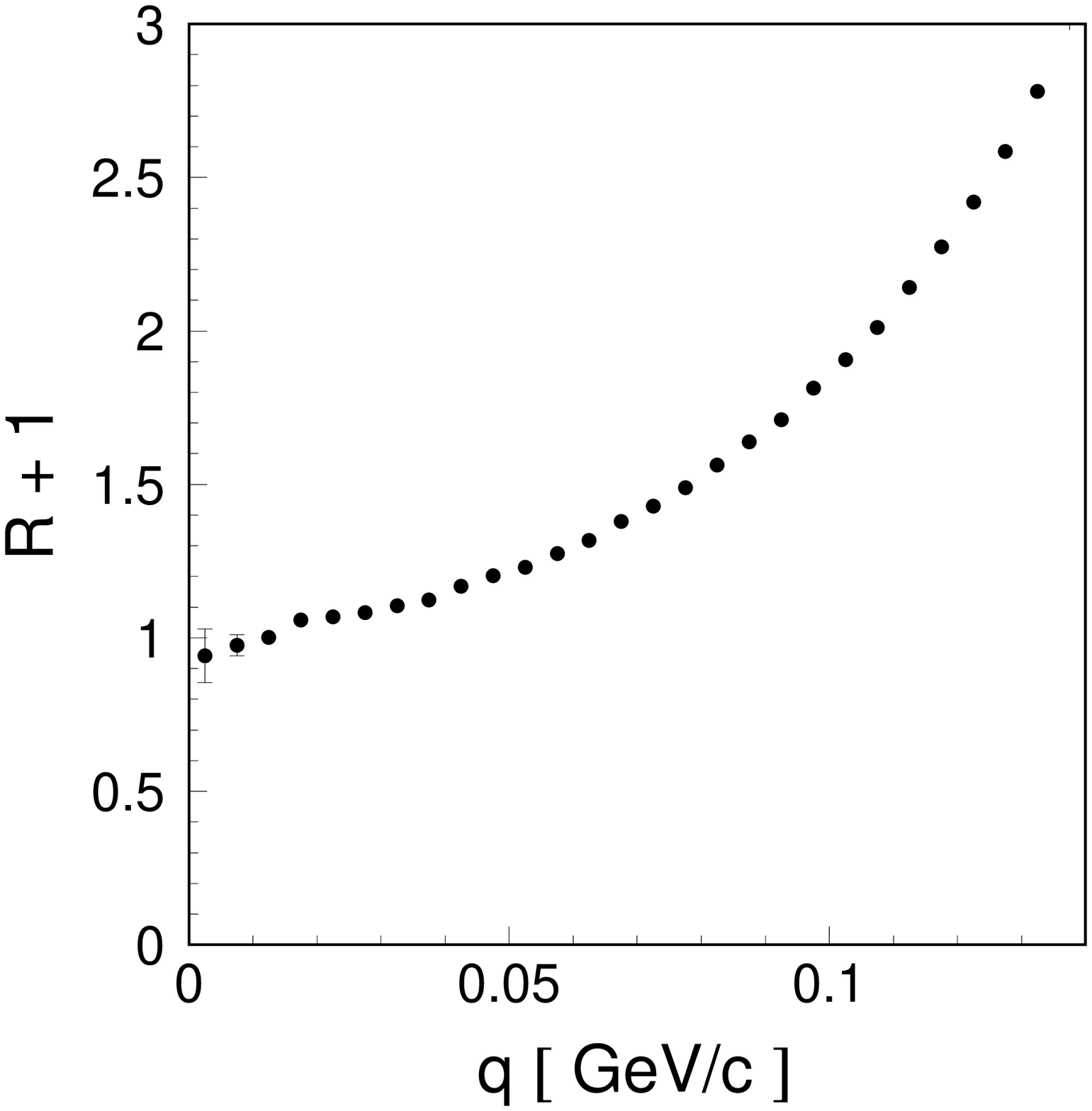}   \hspace{-0.8cm}
  \includegraphics[height=.2\textheight]{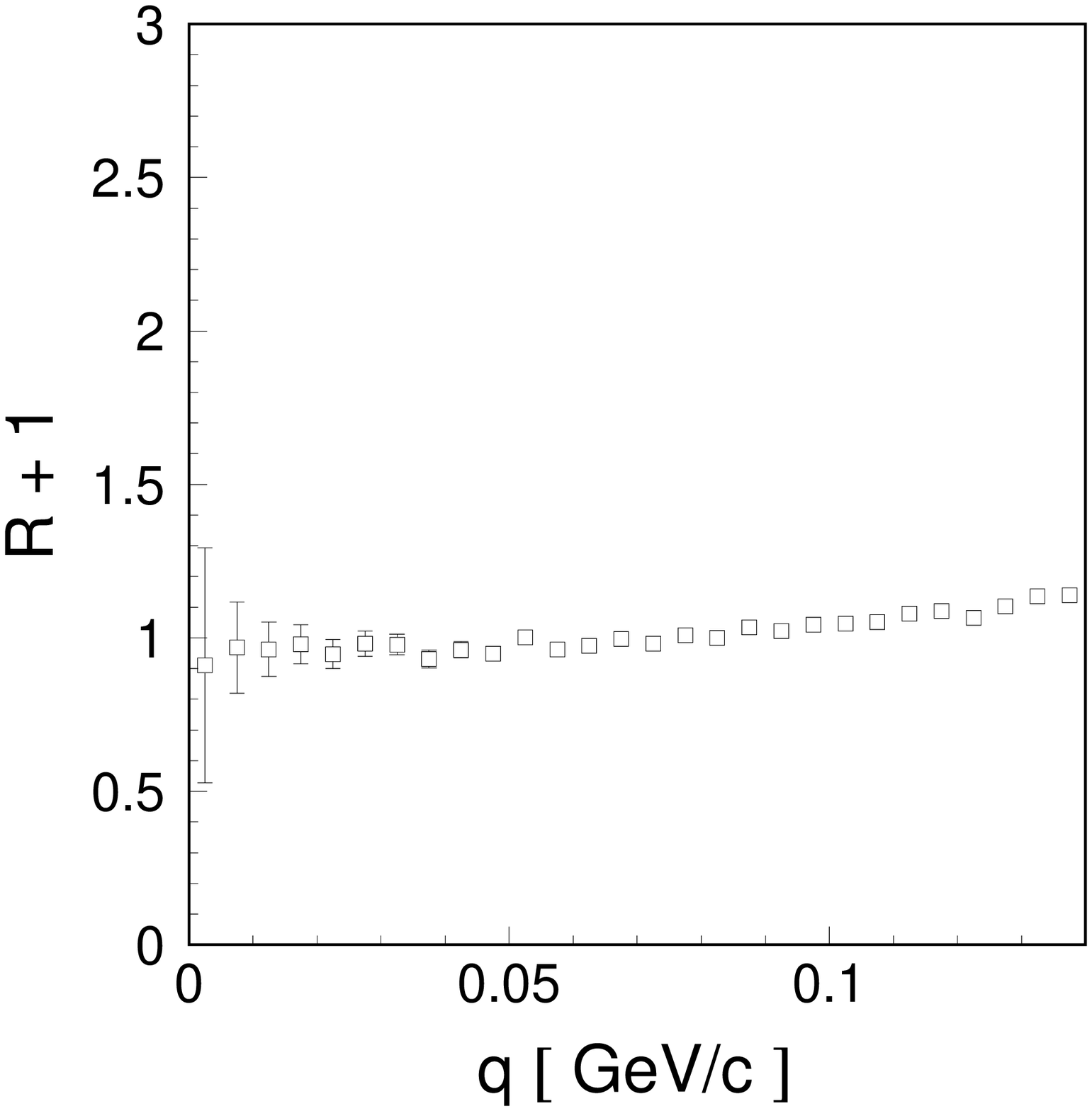}
  \caption{\footnotesize [left panel] Acceptance corrected experimental proton-proton
 correlation functions for the production of the 
  $\eta$ meson (full dots) and multi-pions (open squares). [middle panel]
 The simulated two-proton correlation function for the $\eta$ meson production. 
   [right panel]
 The simulated two-proton correlation function for the $pp \to pp~pions$
                      reactions.}
\label{corr:mcexp}
\end{figure}

The two-proton background-free correlation functions for the 
$pp \to pp\eta$ and $pp \to pp+pions$ 
reactions corrected for the acceptances
are presented in  figure \ref{corr:mcexp}(left). 
As mentioned in the introduction the shape of the obtained 
correlation function reflects not only the
 space-time characteristics of the interaction volume
 but it may also be strongly modified
by the conservation of the energy and momentum and by the final state interaction among the ejectiles.
\begin{figure}[H]
  \includegraphics[height=.28\textheight]{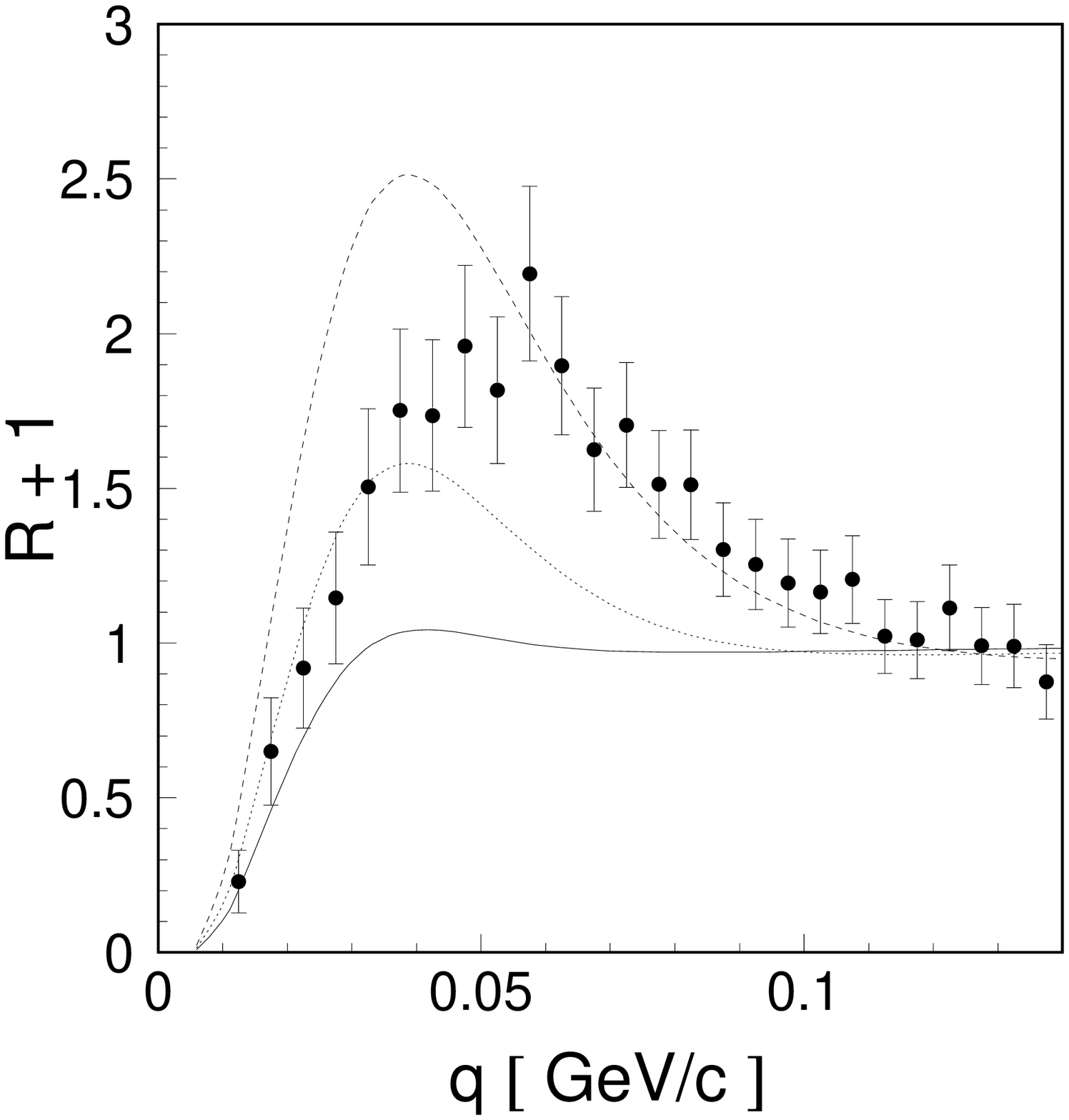}
  \includegraphics[height=.28\textheight]{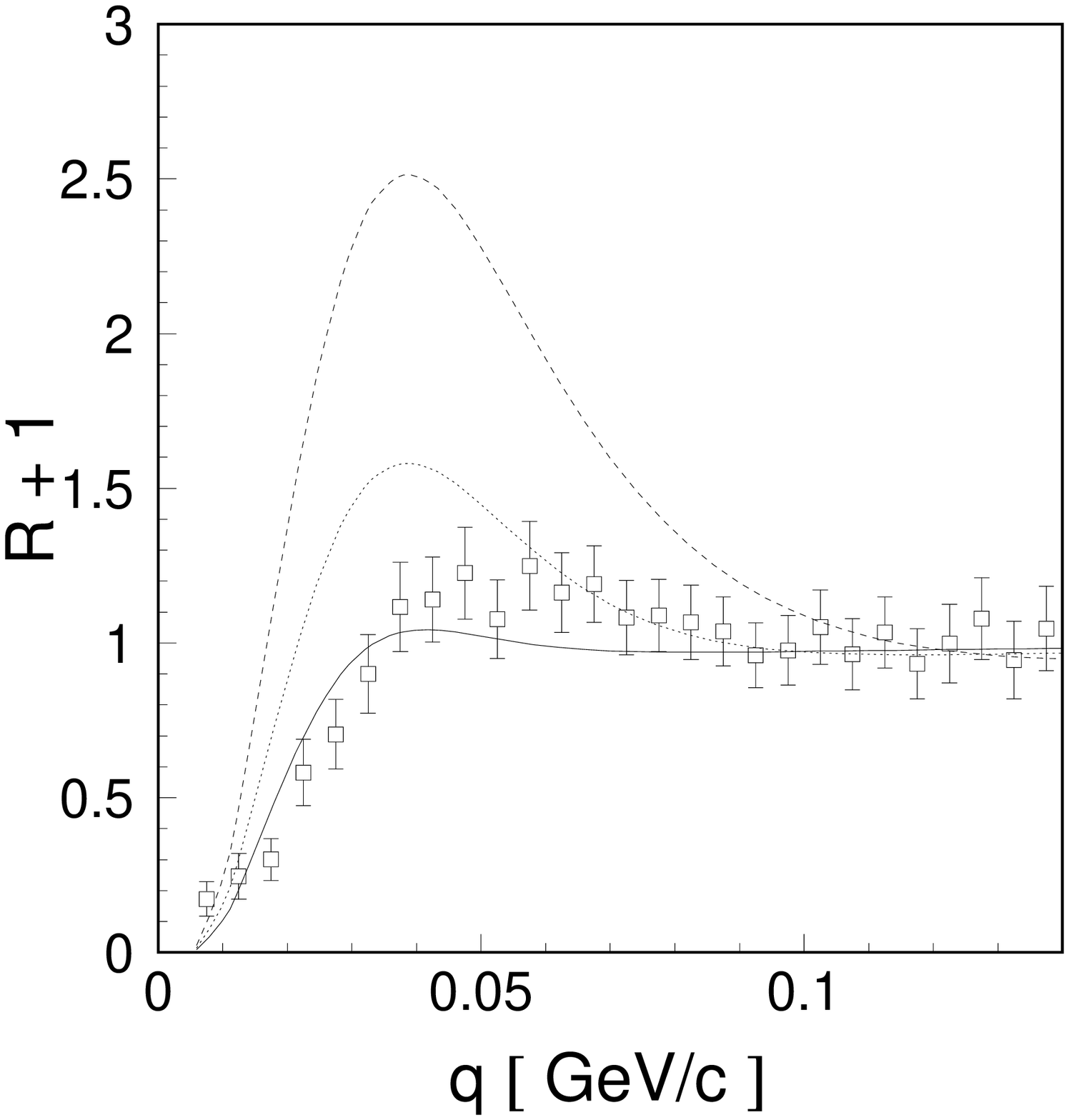}
  \caption{\footnotesize The two-proton
  correlation functions corrected for acceptance and normalized to
  the corresponding correlation function simulated for the point-like source.
  Full dots and open squares represent experimental points
  for the $pp \to pp\eta$ and $pp \to pp+pions$ 
  reaction, respectively. The superimposed lines shows the 
  result of calculations~\cite{deloff}
  for the reaction volume parametrized by a Gaussian 
  with radius $r_{0} = 2.0$ fm (dashed line), $r_{0} = 3.0$ fm (solid line)
  and $r_{0} = 5.0$ fm (dotted line), respectively.}
\label{corr:double}
\end{figure}
In order to estimate the influence of the shape induced by the kinematical bounds we have reconstructed the correlation functions from the data
for both,  the $pp\to pp\eta$ and $pp\to pp+pions$ reaction  
 assuming a point-like source and using a Monte-Carlo simulation.  
The  results of the simulations 
are presented in figure \ref{corr:mcexp} and it is apparent that they differ significantly from the experimental correlation function.
In order to extract from the experimental data the shape of the correlation function free from the influence
of the energy and momentum conservation we had constructed a double ratio:
\begin{equation}
R(q)+1 = C_{exp/MC}~\left(\frac{Y_{exp}(q)}{Y^*_{exp}(q)}/\frac{Y_{MC}(q)}{Y^*_{MC}(q)}\right),
\label{equ:5}
\end{equation}
where $C_{exp/MC}$ denotes the normalization constant, the indices 'exp' and 'MC'  
refer to the experimental 
and simulated samples, respectively. The determined double ratios are presented in figure \ref{corr:double}.
Such procedure is used e.g. by the ALEPH collaboration for studying the correlation of the $\Lambda$ pairs from
the Z decays~\cite{aleph1} or for the studies of correlations in W-pairs decays~\cite{aleph2}.
\par
In order to estimate the size of the emission source the results are compared with
theoretical predictions,  obtained by assuming a 
simultaneous emission of the two protons and derived under the assumption that the final-state interaction between the two detected
particles dominates, while other interactions are negligible.
This is certainly fairly well  satisfied 
in the case of the studied ppX systems.
 The source density was taken to be a Gaussian 
specified by a radius parameter $r_0$ and
further particulars of the calculations
 are presented in reference~\cite{deloff}.
A rough comparison between the theoretical correlation function
and the experimental points indicates that
the effective size of the emission source amounts
 to about 2.4~fm for the $pp\eta$ system
and about 4~fm for the $pp+pions$ system.
 A detailed comparison and the interpretation of results is in progress.

\section*{\normalsize ACKNOWLEDGEMENTS}

We acknowledge the support of the
European Community-Research Infrastructure Activity
under the FP6 programme (Hadron Physics, N4:EtaMesonNet,
RII3-CT-2004-506078), the support
of the Polish Ministry of Science and Higher Education under the grants
No. PB1060/P03/2004/26, 3240/H03/2006/31  and 1202/DFG/2007/03,
and  the support of the German Research Foundation (DFG).

\end{document}